\documentclass[reprint, preprintnumbers, nobibnotes, amsmath,amssymb, prl]{revtex4-2}
\usepackage{amsmath}
\usepackage{braket}
\usepackage{comment}
\usepackage{graphicx}
\usepackage{dcolumn}
\usepackage{bm}
\usepackage{gensymb}
\usepackage{url}
\usepackage{color}
\usepackage{hyperref}
\begin{document}

\preprint{WSU-HEP-2504}
\title{Gedanken Experiments of Entanglement in Particle Physics: \\ Interactions, Operators and Bell-Type Inequalities in Flavor Space}

\author{Corbin Pacheco}
\email{cpacheco@wayne.edu}
\author{Nausheen R.~Shah}
 \email{nausheen.shah@wayne.edu}
\affiliation{
Department of Physics \& Astronomy \\
Wayne State University\\
Detroit, MI 48201, 
}

\date{\today}

\begin{abstract}
We propose Bell-type inequalities by constructing flavor operators associated with mass identification, flavor change, and charged-current weak mixing which arise from fundamental interactions in the Standard Model. We treat these interactions as measurement settings in a Gedanken experiment. For entangled two-particle states, the measurement outcomes of these flavor operators exhibit correlations that violate a Bell-type inequality under the stated assumptions, showing that our result probes contextuality and incompatibility of interaction-defined observables. We discuss how the predicted correlation patterns may be probed with experimental data, clarifying the scope and limitations of Bell-type reasoning in particle physics.
\end{abstract}
\maketitle
\section{\label{sec:level1}Introduction:}
The Standard Model (SM) of particle physics is based on the gauge symmetry $SU(3)_C \times SU(2)_L \times U(1)_Y$, and contains three generations of fermions with seemingly arbitrary masses and mixings. This interplay between symmetry and apparent arbitrariness in the SM invites some scrutiny. Concepts such as entanglement and entropy, central to quantum information theory, offer a promising lens through which to examine this structure. Recent work has explored theoretical interactions as sources of entanglement and entropy, as well as the use of quantum-information-theoretic tools to formulate constraining principles within the SM and beyond, see for e.g. Refs.~\cite{Carena:2025wyh, Kowalska:2025qmf, Liu:2025bgw, Liu:2025qfl, Hu:2025jne, Busoni:2025dns, Kowalska_2025, Siwach_2025, ThalerTrifinopoulos2025,  CarenaLowWagner2024, SinhaZahed2023, Fedida_2023, Liu:2022grf, Beane_2021, BeaneKaplan2019, Cervera_Lierta_2017}.

While there have been significant investigations of flavor entanglement in meson systems, most collider experiments investigating entanglement have targeted spin correlations in particle pair production as proposed by Bell~\cite{Bell1964}, see for e.g. Refs.~\cite{ATLAS:2023spin,CMS:2025anw,Afik:2024,Low:2025aqq,CMS2024TopEntanglement,HanLowSu2025,HanZhouEtAl2023,Dong:2023xiw}. We emphasize that Bell-type inequalities provide a general  criterion for distinguishing classical from quantum correlations in bipartite systems, and  test entanglement through correlations between binary outcomes. While the original proposal was concerning spin~\cite{Bell1964}, their defining feature is that they rely only on minimal, theory-independent assumptions and are agnostic to the microscopic realization of the observables involved~\cite{JFClauser_1978,Abramsky:2012dlv}.  Crucially,  Bell-type tests do not depend on the actual numerical value of an observable, but only on the existence of two distinguishable outcomes, often labeled 
$+1$ and $-1$.  

In collider experiments, we do not measure spin directly in the same way as in tabletop experiments. Instead, we predominantly measure momenta and reconstruct mass eigenstates~\cite{CMS:2025anw, CMS2024TopEntanglement, Abel:2025skj,Bechtle:2025ugc, Li:2024luk, ABEL1992304}. On the other hand, collider experiments are fundamentally counting experiments, making them naturally compatible with the logic of Bell-type inequalities~\cite{JFClauser_1978, Abramsky:2012dlv, AbramskyBrandenburger2011, BrunnerReview2014}, provided suitable binary operators/measurements can be identified. In particle physics, in addition to spin, internal quantum numbers such as flavor naturally define effective two-level systems, but the corresponding observables are accessed through distinct SM interaction channels rather than tunable detector orientations. This motivates reformulating Bell-type reasoning in a setting where measurement settings are defined by physical interactions.

 We propose using fundamental SM interactions as probes of entanglement in particle physics. Such interactions will be shown to be completely analogous to spin operators in quantum mechanics, performing binary measurements of inherent particle quantum numbers. 

We construct dichotomic operators associated with 
flavor~(mass identification, flavor change, and charged-current weak mixing), and demonstrate their mutual incompatibility. Applying this framework, we show that the algebraic structure of these operators alone suffices to generate Bell-type inequality violations for entangled two-particle states, independent of kinematic correlations of decay products.  Possible experimental validation of such violations is also discussed.

\section{Binary Observables and Bell-Type Reasoning}\label{sec:Bell}

Bell in his seminal paper~\cite{Bell1964} discussed 
quantum versus classical correlations for the spin singlet state. Here, for completeness, we present his argument for both correlated and anti-correlated states. Consider the maximally entangled spin states,
\begin{align}
    |\Psi\rangle_+
&=
\frac{1}{\sqrt{2}}
\left(
|\uparrow\uparrow\rangle
+
|\downarrow\downarrow\rangle
\right)\;, \quad\textrm{or}\label{eq:corr}    \\
    |\Psi\rangle_-
&=
\frac{1}{\sqrt{2}}
\left(
|\uparrow\downarrow\rangle
-
|\downarrow\uparrow\rangle
\right)\;.~ \label{eq:uncorr}
\end{align}

Assume a local hidden variable theory (LHVT) characterized by a hidden variable $\lambda$, distributed according to a density $\rho(\lambda)$. Given locality and realism as encoded in the LHVT, if Alice measures the first qubit and Bob measures the second qubit, measurement outcomes are determined by functions
$A(a,\lambda)$ and $~ B(b,\lambda)$ respectively,
corresponding to measurement choices $a$ and $b$, with binary values
\begin{equation}\label{eq:binary}
A(a,\lambda)=\pm1, \qquad B(b,\lambda)=\pm1.
\end{equation}
The correlation in each measurement is given by,
\begin{equation}
    A(a,\lambda)  B(a,\lambda) = \pm 1\;,
\end{equation}
where the $\pm$ refers to the correlated,~Eq.~(\ref{eq:corr}) or anti-correlated state,~Eq.~(\ref{eq:uncorr}), respectively. 

After a series of measurements performed by Alice and Bob on identically prepared initial states, the expectation value of the product of their measurements predicted by the LHVT is
\begin{equation}
E(a,b)
=
\int d\lambda\, \rho(\lambda)\, A(a,\lambda)\,B(b,\lambda).
\end{equation}
Given a minimum of three measurement choices $a$, $b$ and $c$, Bell's inequality may then be written in the form
\begin{equation}
\bigl|E(a,b) - E(a,c)\bigr|
\;\le\;
1 \pm E(b,c)\;,
\label{eq:bell_original}
\end{equation}
where the + (\,-\,) on the right hand side corresponds to the initial state being anti-correlated (correlated). This bound, often referred to as Bell's Inequality, limits the maximum possible correlation in measurements that could be obtained in a classical system. 

At the operator level, Bell-type inequalities test whether a LHVT can consistently assign predetermined values to incompatible observables so as to reproduce all correlators, independent of the details of any experimental implementation~\cite{Bell1964,JFClauser_1978,Abramsky:2012dlv, Fine:1982,AbramskyBrandenburger2011,BrunnerReview2014}. 
Further, the analysis above is independent of the physical realization of the observables and only depends on the binary nature of the system, Eq.~(\ref{eq:binary}). Hence, any 2-state system where we can identify operators/measurements corresponding to the spin {\it identification} ($\hat{S}_z$),  spin {\it flip} ($\hat{S}_x$) and  spin {\it mixing} ($\hat{S}_\theta$) operators may be analyzed in exactly the same framework. Therefore, to test Bell-type inequalities at colliders we need only identify three independent binary measurement outcomes due to SM particle physics interactions and put them in correspondence with the spin measurements listed above.

\section{Flavor as a Fundamental 2-Level System in Particle Physics}
\label{sec:single}

There is a growing body of literature arguing that spin-based Bell tests are challenging in collider environments due to the lack of controllable spin analyzers, rapidly varying reference frames, and dependence on momentum reconstruction~\cite{Abel:2025skj,Bechtle:2025ugc, Li:2024luk, ABEL1992304}. However, the SM  contains other intrinsic two-level quantum degrees of freedom at the single-particle level; among these is flavor, which we discuss below.

We construct explicit operators associated with binary observables in particle-physics processes in analogy with measurement angles in the canonical Bell experiments. These operators are defined through the structure of weak interactions and decay channels, coupled with direct projective mass measurements, and are therefore interpreted as effective observables at the operator level. We emphasize that all observables and correlations are constructed from intrinsic SM interactions. The appearance of a Pauli-algebra structure reflects the  $SU(2)$ operator algebra of the restricted two-generation flavor subspace rather than an imposed toy model. This construction provides the ingredients needed for the Bell-type analysis developed in the following section.

\subsection{Fermion Generations: Flavor and Mass}
\label{sec:flavor}

 Flavor~(or generation) is not a conserved quantum number of the weak interactions in the SM. While the SM contains three generations of fermions, we restrict our attention here to the first two generations. The restriction to two generations is not an approximation to the SM but a deliberate projection onto a two-dimensional subspace. Hence, effective operators acting on a restricted flavor Hilbert space can be defined in a precise analogy with spin operators in quantum mechanics. This mirrors the use of polarization qubits in quantum optics~\cite{FabreTreps2020,romero2025photonicquantumcomputing, NielsenChuang2012} and provides the minimal Hilbert space required for constructing Bell-type inequalities.

In the SM, flavor eigenstates $|f_a\rangle$ diagonalize the interaction basis, while mass eigenstates $|m_a\rangle$ diagonalize propagation in space-time. 
Detectors identify particles through invariant mass, lifetime, and decay kinematics, all of which correspond to the mass basis. Thus, while interactions are defined in flavor space, measurements project onto mass eigenstates.  Restricting attention to two flavors, the single-particle mass and flavor Hilbert spaces are spanned by $\left\{ |m_1\rangle, |m_2\rangle \right\}$, and $\left\{ |f_1\rangle, |f_2\rangle \right\}$, denoting the mass eigenstates and the eigenstates of the charged weak interactions respectively. The two bases are related by a unitary transformation,
\begin{equation}\label{eq:fmass}
|f_i\rangle = \sum_a U_{ia} |m_a\rangle,
\end{equation}
where $U$  is the appropriate  mixing matrix (e.g. CKM or PMNS). 

In the following, for concreteness, we consider the quark sector. Hence, the relevant rotation matrix for our chosen bases is the Cabibbo matrix, defined in terms of the Cabibbo angle $\theta_c$, 
\begin{equation}
\hat{U}=c_{\theta_c}\mathbb{I}+i s_{\theta_c}\hat{\sigma}_y \;,
\end{equation}
where $\mathbb{I}$ is the identity matrix and $c_{\theta_c}$ and $s_{\theta_c}$ denote $\cos\theta_c$ and $\sin\theta_c$ respectively. We assume that the up quark sector is aligned with the mass basis and the down quark sector~$\{d,s\}$ contains all the flavor violation. However, we do not distinguish between particle/anti-particle or up  versus down  type fermions in each generation for defining our Hilbert space. Hence, for example, $|m_1\rangle$ labels any of the first generation quarks: $\{u, \bar{u}, d, \bar{d}\}$~\footnote{The different masses and charges are relevant for detector measurements. This is no different from considering the Hilbert space for spins as $\{\uparrow,~\downarrow\}$ regardless of the particle being an electron or a positron for example, even though that may play a role in the categorization of the final measurements.}.  

We now construct {\it flavor operators} in the SM in analogy with the spin operators of quantum mechanics.
We begin by first defining the Hermitian mass identification operator,
\begin{equation}
\hat{F}_{ID}=
\hat{\sigma}_z= |m_1\rangle\langle m_1| - |m_2\rangle\langle m_2|,
\end{equation}
which has eigenvalues $\pm 1$,
\begin{equation}
\hat{F}_{ID} |m_{1,2}\rangle = \pm |m_{1,2}\rangle\;, 
\end{equation}
corresponding to whether the  mass measured in a detector corresponds to a fermion belonging to the first or the second generation.

Once $\hat{F}_{ID}$ is defined, a complete operator basis for our flavor states is given by,
\begin{align}
\hat{F}_x &= |m_1\rangle\langle m_2| + |m_2\rangle\langle m_1|, \label{eq:Fx}\\
\hat{F}_y &= -i|m_1\rangle\langle m_2| + i|m_2\rangle\langle m_1|. \label{eq:Fy}
\end{align}
These operators satisfy the $SU(2)$ algebra,
\begin{equation}
[\hat{F}_i, \hat{F}_j] = 2i \epsilon_{ijk} \hat{F}_k,
\end{equation}
where $\{i,j,k\}$ corresponds to $\{x,y,ID\}$ above, and thus generate rotations in flavor space.
The existence of non-commuting flavor operators is essential: Bell-type inequality violations require multiple measurement settings corresponding to incompatible observables.

Different {\it measurement settings} correspond to measurements along a rotated flavor basis,
\begin{equation}
\hat{F}(\theta) = \cos\theta\, \hat{F}_{ID} + \sin\theta\, \hat{F}_\perp,
\end{equation}
where $\hat{F}_\perp$ is an orthogonal flavor operator within the same two-dimensional space. However, unlike spin systems, where one accomplishes this by changing the angle of the polarizer or magnetic field used in the measurement, a flavor rotated measurement in the collider environment is obtained via dynamic SM interactions followed by detector measurements. 

Following this logic, the observation of a mass eigenstate after a charged weak interaction~(scattering with a $W$) may be represented by the flavor mixing operator,  
\begin{eqnarray}
\hat{F}_W &=& \hat{U}^{\dagger}\hat{F}_{ID}\hat{U} = (c_{\theta_c}^2-s_{\theta_c}^2) \hat{\sigma}_z + 2c_{\theta_c}s_{\theta_c}\hat{\sigma}_x\;, \\
&=& \hat{F}(2\theta_c)\;.
\end{eqnarray}

In addition to $\hat F_{ID}$ and $\hat F_W$, corresponding to $\hat S_z$ and $\hat S_{2\theta_c}$ respectively, we also need
an operator corresponding to a third direction in flavor space such as $\hat S_x$. Such an operator in flavor space would correspond to a flavor changing interaction. While flavor changing neutral currents~(FCNCs) are forbidden at tree-level in the SM, they are present at the loop level mediated by the charged weak interactions and are responsible for phenomena such as meson-anti-meson mixing. Here we will focus on the kaon system. 

Given a $|d\rangle$ quark~($|m_1\rangle$), strong interactions with the background can produce the neutral kaon~$|K^0\rangle = |d\bar{s}\rangle$. Similarly, an $|s\rangle$ quark~($|m_2\rangle$) may produce a neutral anti-kaon~$|\overline{K}^0\rangle = |s\bar{d}\rangle$. We denote this process via the effective kaon map
$\hat F_K$,
\begin{equation}\label{eq:Kk}
    \hat F_K = | K^0\rangle\langle d| + |\overline{K}^0\rangle\langle s|\;.
\end{equation}
The states that undergo distinctively different weak hadronic decays and have definite observed masses and lifetimes, $|K_S\rangle$ and $|K_L\rangle$ are  superpositions of the two neutral kaons~\cite{Christenson1964}, 
\begin{eqnarray}
    |K_S\rangle&=&\frac{1}{\sqrt{2}}\left(|K^0\rangle + |\overline{K}\rangle^0 \right) \;,\label{eq:KS}\\
    |K_L\rangle&=&\frac{1}{\sqrt{2}}\left(|K^0\rangle - |\overline{K}^0\rangle \right) \label{eq:KL}\;.
\end{eqnarray}
Interestingly, $|K_S\rangle$ and $|K_L\rangle$
are precisely the eigenstates of the operator $\hat \sigma_x$
if $|K^0\rangle$ and $|\overline{K}^0\rangle$ are associated 
with the eigenstates of the $\hat \sigma_z$ operator. For simplicity, CP violation effects are neglected in this idealized construction~\footnote{We neglect CP violation for clarity. Including the small CP violating parameter $\epsilon_K$ modifies the idealized eigenstates $|K_S\rangle$ and $|K_L\rangle$. However, this does not qualitatively alter our construction and can be incorporated straightforwardly if required.}.
The existence of $|K_S\rangle$ and $|K_L\rangle$ thus motivates an effective flavor-decay observable $\hat F_D$ with binary outcomes $\pm1$, and eigenstates,
\begin{equation}
    \hat{F}_{D} ~|K_S\rangle = +|K_S\rangle, \quad     \hat{F}_{D} ~|K_L\rangle = -|K_L\rangle \;,
\end{equation}
where the two eigenvalues correspond to the observation of either the $2\pi=+1$ or $3\pi=-1$ decay mode. 

Explicitly, the operator $\hat{F}_{D}$ is a tensor operator acting on 2-particle states,
\begin{equation}
    \hat{F}_{D} |m_i \overline{m}_j\rangle \equiv \hat{F}_x \otimes \hat{F}_x |m_i \overline{m}_j\rangle = |m_j \overline{m}_i\rangle \;,
\end{equation}
with $\hat{F}_x$ defined as in Eq.~(\ref{eq:Fx}). As mentioned above, $\hat{F}_x$ does not exist in the SM at tree-level for single particle states. Nevertheless, the operator $\hat F_D$ suffices for our purposes, as it permits the definition of an effective flavor--flip operator,
\begin{equation}\label{eq:Ff}
\hat F_F \equiv \hat F_K^{\dagger}\,\hat F_D\,\hat F_K ,
\end{equation}
using the correlation between neutral--kaon states and the mass eigenstates $|m_i\rangle$, Eq.~(\ref{eq:Kk}). Here $\hat{F}_K^\dagger$ is introduced only as part of an operator-level construction of the effective correlator in the underlying flavor space. It should not be interpreted as a physical inverse of the hadronization process, which is intrinsically non-unitary in the open-system description.

 The operators $\hat{F}_K$,  $\hat{F}_D$ and $\hat{F}_F$ should be understood as effective binary observables defined by experimentally accessible decay channels. Because hadronization is an open quantum process, these mappings do not correspond to unitary transformations on a closed two-level system, but are most appropriately interpreted as effective binary positive operator valued measures~(POVMs)~\cite{NielsenChuang2012,Smolinski:2015dpa, Bernabeu:2012au, Benatti:1999gw}. The two-level structure used here is an effective description at the level of experimentally accessible observables, and does not imply that the kaon system is a closed qubit. This does not affect the operator-algebraic structure underlying the Bell-type inequality being constructed here.

Finally, while algebraically similar $SU(2)$ structures have been used in meson systems, including kaons~\cite{Chen:2024drt, Bertlmann:2004yg, Bertlmann:2001sk}, the present approach differs both in construction and interpretation. In particular, we consider entangled quarks rather than mesons. Moreover, interaction channels are identified as measurement operators, yielding a closed set of incompatible observables directly tied to physical processes. The resulting Bell-type constraint will be expressed in terms of a measured SM parameter (the Cabibbo angle) and does not rely on oscillatory dynamics such as in $K^0$--$\bar{K}^0$ mixing, but instead on the incompatibility of interaction-defined observables.

The measurement outcomes of the set of the flavor operators $\{\hat{F}_{ID},\hat{F}_{F},\hat{F}_{W}\}$   on the mass eigenstates $\{|m_1\rangle,~|m_2\rangle\}$, exactly mimic  the outcomes of the spin operators $\{\hat{S}_z, \hat{S}_x , \hat{S}_\theta\}$ on the spin states $\{|\uparrow_z\rangle,~|\downarrow_z\rangle\}$. Hence we have identified three (SM interactions + mass measurements) flavor operators corresponding to identification, mixing and flip in the spin case, the minimal requirement for the formulation of Bell-type inequalities. We are now in the position of directly applying these flavor operators to an entangled flavor state to test Bell-type inequalities for such a system, Eq.~(\ref{eq:bell_original}).

\section{Two-Particle Wave Functions}
\label{sec:2part_wavefunctions}

The minimal setting in which quantum entanglement can arise is a composite system consisting of two subsystems with internal degrees of freedom. Flavor is a distinct quantum number in the SM, correlated non-trivially by different interactions. These interactions then provide a natural mechanism by which  entanglement is generated, alleviating the need for external entangling operations. 

Below we construct an explicit example of Bell-type inequalities  using the flavor formalism developed in the previous section. Rather than relying on actively chosen detector settings, the observables are defined through interaction contexts, i.e. the physical processes by which the system is probed. This formulation allows Bell-type inequalities to be used as operator-level diagnostics of local hidden-variable descriptions, while keeping the assumptions and limitations of such constructions explicit. 
The Bell-type inequality derived here relies on the assumption that each pair-wise measurement, collected across different interaction contexts, originates from the same underlying quantum state prepared at production.

\subsection{Entangled Flavor Bell State}
Consider the production of fermion--antifermion pairs giving rise to the entangled flavor state,
\begin{equation}\label{chi}
    |\chi\rangle =\frac{1}{\sqrt{2}}\left( |d\bar{d}\rangle +|s\bar{s}\rangle \right) \;,
\end{equation}
where each particle in the pair moves freely in the opposite direction, completely analogous to the entangled correlated spin state of Eq.~(\ref{eq:corr}).

In SM processes such as $Z$-boson decays, an entangled wave function consisting of all possible final states is naturally created. The state $|\chi\rangle$ should therefore be understood as an idealized two-generation flavor projection of such a wave function, capturing the relevant correlation structure for the Gedanken analysis. Realistic production/hadronization reduce its fidelity but do not alter the operator-algebraic structure of the analysis.

\begin{table*}[t]
\centering
\begin{ruledtabular}
\begin{tabular}{ccccc}
$(x,y)$
& $(\hat F^A \otimes\hat F^B)$
& (Alice, Bob)
& Weighted Probability Joint Outcomes 
& $E(x,y)$ \\ \hline \hline

$(a,b)$
& $(\hat F_F\otimes\hat F_{ID})$ 
& $(K_S,\bar d),(K_L,\bar d)$ 
& $\tfrac14(+1)(+1)+\tfrac14(-1)(+1)$
& 
\\

& 
& $(K_S,\bar s),(K_L,\bar s)$ 
& $\tfrac14(+1)(-1)+\tfrac14(-1)(-1)$
& 
\\

&&&& $0$ \\
\hline

$(a,c)$
& $(\hat F_F\otimes\hat F_W)$
&$(K_S,\bar u)$,
$(K_L,\bar u)$  
& $(+1)(+1)\tfrac14(1+s_{2\theta_c})+(-1)(+1)\tfrac14(1-s_{2\theta_c})$ 
& 
\\

& 
&$(K_S,\bar c)$,
$(K_L,\bar c)$  
& $(+1)(-1)\tfrac14(1+s_{2\theta_c})+(-1)(-1)\tfrac14(1-s_{2\theta_c})$ 
& 
\\

&&&& $s_{2\theta_c}$  \\
\hline

$(b,c)$
& $(\hat F_{ID}\otimes\hat F_W)$
& $(d,\bar u)$ ,
$(d,\bar c)$  
& $(+1)(+1)\tfrac12c^2_{\theta_c}+(+1)(-1)\tfrac12s^2_{\theta_c}$
& \\

& 
& $(s,\bar u)$ ,
$(s,\bar c)$  
& $(-1)(+1)\tfrac12s^2_{\theta_c}+(-1)(-1)\tfrac12c^2_{\theta_c}$
& \\

&&&& $c_{2\theta_c}$
\end{tabular}
\end{ruledtabular}
\caption{
Joint measurement outcomes and correlations for the entangled
state $|\chi\rangle=\tfrac{1}{\sqrt2}(|d\bar d\rangle+|s\bar s\rangle)$. Probabilities and expectation values follow directly from Eqs.~(\ref{eq:Eab})-~(\ref{eq:Ebc}) as explained in the text. Outcomes for $A \leftrightarrow B$ can be obtained by exchanging $K_L\leftrightarrow K_S$ and $f\leftrightarrow \bar{f}$.
}
\label{tab:flavor_corr}
\end{table*}

We  write the measurements by Alice and Bob as the tensor operator $(\hat{F}^A\otimes\hat{F}^B)$ acting on the state $|\chi\rangle$ where $A$ acts on the first qubit and $B$ on the second qubit respectively. Each experimental run realizes a definite pair of interaction-defined measurement settings $(a,b)$ acting simultaneously on the two subsystems, and yields a corresponding pair of outcomes from which the correlator 
$E(a,b)$ is defined operationally. Different runs give rise to different pairs of outcomes. The expectation values of the three combinations 
\begin{eqnarray}
E(a,b)&=\langle\hat{\sigma}_{x}^A\otimes\hat{\sigma}_{z}^B\rangle_{\psi_+}=\langle\hat{F}_{F}^A\otimes\hat{F}_{ID}^B\rangle_\chi&=0, \label{eq:Eab} \\
E(a,c)&=   \langle\hat{\sigma}_{x}^A\otimes\hat{\sigma}_{2\theta}^B\rangle_{\psi_+}= \langle\hat{F}_F^A\otimes\hat{F}_W^B\rangle_\chi&=s_{2\theta_c}, \label{eq:Eac}    \\
   E(b,c)&=\langle\hat{\sigma}_{z}^A\otimes\hat{\sigma}_{2\theta}^B\rangle_{\psi_+}=\langle\hat{F}_{ID}^A\otimes\hat{F}_W^B\rangle_\chi&=c_{2\theta_c},\label{eq:Ebc}
\end{eqnarray} 
can be used to test the LHVT limit for the correlated system~Eq.~(\ref{eq:bell_original}), 
\begin{equation}
\bigl|E(a,b) - E(a,c)\bigr|
\;\le\;
1 - E(b,c)\;.
\label{eq:bell_corr}
\end{equation}
Using Eqs.~(\ref{eq:Eab})-~(\ref{eq:Ebc}) for the flavor system, this corresponds to,
\begin{equation}
    |s_{2\theta_c}|+c_{2\theta_c} \le 1\;.
\end{equation}
The experimentally measured value of $2\theta_c\sim 26\degree$~\cite{PDG2024PRD} gives,
\begin{equation}\label{result}
    \sim 0.44 + 0.90 = 1.34 \nleq 1\;,
\end{equation}
in clear violation of Eq.~(\ref{eq:bell_corr}) in the idealized settings considered here.

It is instructive to detail the  detector observations expected in each of the cases above, summarized in Table~\ref{tab:flavor_corr}. We will address first the expectation values where the mass of at least one of the particles is directly measured via $\hat{F}_{ID}$. Hence for $E(a,b)$, Eq.~(\ref{eq:Eab}), when a $\bar{d}$ quark is observed by $B$, an equal number of $K_S$ and $K_L$ decays will be observed by $A$, giving rise to the joint expectation value of 0. Similarly for an $\bar{s}$ measurement by $B$. Hence, in agreement with Eq.~(\ref{eq:Eab}), 0 is obtained for the sum. For $E(b,c)$, Eq.~(\ref{eq:Ebc}), when a $d$ quark is observed by $A$, $c_{\theta_c}^2$ of the time, $\bar{u}$ will be observed by $B$, and $s_{\theta_c}^2$ of the time, a $\bar{c}$ will be observed. Inverse results will be observed when an $s$ quark is observed by $A$. Hence, the total correlation expectation value for this set of measurements will be 
\[\hspace{-0.1cm}\frac{1}{2}\Big[(+)\Big((+)c_{\theta_c}^2+(-)s_{\theta_c}^2\Big)+(-)\Big((-)c_{\theta_c}^2+(+)s_{\theta_c}^2\Big)\Big]=c_{2\theta_c}\]
as expected. Finally, for $E(a,c)$, Eq.~(\ref{eq:Eac}), we look at the mass measurement outcomes for the state $|\chi\rangle$ {\it after} scattering with the $W$ interaction via the rotation operator for the one qubit and the kaon creation operator for the other, 
\begin{eqnarray}
   |\chi'\rangle&=&\hat{F}_K^A \otimes \hat{U}^B |\chi\rangle = \nonumber \\
   && \frac{1}{\sqrt{2}}\left[ \frac{1}{\sqrt{2}}(|K_S\rangle+|K_L\rangle )\otimes (c_{\theta_c}|\bar{u}\rangle-s_{\theta_c}|\bar{c}\rangle) \right.\nonumber\\
   && \left. + \frac{1}{\sqrt{2}}(|K_S\rangle-|K_L\rangle )\otimes (c_{\theta_c}|\bar{c}\rangle+s_{\theta_c}|\bar{u}\rangle)
   \right]. \label{eq:chip}
\end{eqnarray}
Then, $E(a,c)=\langle\hat{F}_F^A\otimes\hat{F}_W^B\rangle_\chi  = \langle\hat{F}_D^A\otimes\hat{F}_{ID}^B\rangle_{\chi'}$ which  can be read off from the above:  The joint probability for the observation of a $\bar{u}$ by $B$ together with $K_S$ decays by $A$ is $\frac{1}{4}(c_{\theta_c}+s_{\theta_c})^2=\frac{1}{4}(1+s_{2\theta_c})$, and with $K_L$ decays observed is $\frac{1}{4}(c_{\theta_c}-s_{\theta_c})^2=\frac{1}{4}(1-s_{2\theta_c})$. If instead a $\bar{c}$ is observed by $B$, the joint probabilities with the respective kaon decay observations are inverted. Hence, the correlation between the two measurements is given by
$[(+)(+)\frac{1}{4}(1+s_{2\theta_c})+(-)(-)\frac{1}{4}(1+s_{2\theta_c})+ (+)(-)\frac{1}{4}(1-s_{2\theta_c}) +(-)(+)\frac{1}{4}(1-s_{2\theta_c})]=s_{2\theta_c}$. This is again in agreement with the naive expectation value arising from the standard spin algebra presented in Eq.~(\ref{eq:Eac}).

In this framework, measurement contexts are defined by physical interactions (mass identification, weak mixing, and flavor change), rather than freely tunable analyzer settings. We emphasize that the specific joint outcome identifies which interaction channels probed the two subsystems in that run, thereby determining the effective measurement context. Because the effective measurement context is inferred from observed event signatures, this framework does not enforce strict measurement independence in the Bell sense, but instead tests the consistency of LHVT under the stated idealizations. Hence our result, Eq.~(\ref{result}) should be interpreted as a Bell-type consistency condition for the existence of a joint assignment of predetermined values to the observables~\cite{Fine:1982}, rather than as a direct test of spacetime locality. Its violation implies the incompatibility of a noncontextual hidden-variable description for these observables.

\subsection{Experimental Validation}

The purpose of this section is to  illustrate how the operator-level correlations derived above may be connected to experimentally accessible signatures. The Bell-type inequality constructed here should be interpreted as a consistency condition on correlations among interaction-defined observables within the idealized framework considered above. Experimentally,  the essential requirement is access to the distinct interaction-defined observables entering the correlators -- mass identification, flavor change, and charged-current mixing -- on ensembles prepared in the same underlying quantum state(such as an ensemble of events originating from $Z$-boson decays). The correlations described in the previous section are obtained by pairwise observations in which both subsystems are measured in the same event, with the interaction channel inferred from the observed decay and mass signatures. 

We emphasize that present experiments may not realize all assumptions of the idealized setting choices envisioned here; nevertheless, the Bell-type inequality proposed here provides a consistency condition that any LHVT description of flavor observables must satisfy. Agreement with our predicted correlation pattern would support the operator-level description and entanglement structure assumed in this work, while significant deviations would indicate either experimental limitations or the need for an alternative description of flavor correlations beyond the class of noncontextual hidden-variable descriptions considered here.

Finally, we note that while we have constructed an explicit realization of Bell-type inequalities  for the first and second generation of quarks, an analogous analysis may be undertaken for the first and second lepton generations, with a straightforward replacement of relevant parameters: $f_1: \{u, \bar{u}, d,\bar{d}\}\rightarrow \{e^-,e^+,\nu_e,\bar{\nu}_e \}$, etc. However, even at the Gedanken level, experimentally probing the lepton sector for Bell-type inequalities may be more challenging than for the quarks: Neutrinos are notoriously difficult to detect; muonium-antimuonium oscillations (while possible in the SM due to non-zero neutrino masses) have yet to be observed; and neutrinos don't form bound states.

\section{Discussion}\label{sec:disc}

The Standard Model stands as one of the most successful theories in physics, yet its unnatural features continue to challenge our understanding. Exploring ideas such as entanglement, operators, and information-theoretic concepts may provide new ways of thinking about long-standing puzzles. Entanglement in particle physics remains largely unexplored, and the collider environment offers a unique laboratory for testing quantum correlations at the highest accessible energies.

The framework we have presented demonstrates that Bell-type inequalities can be meaningfully formulated in particle physics without direct spin measurements. Flavor provides a natural binary observable intrinsic to the SM, even though measurements are performed on mass eigenstates.

The Bell-type inequality proposed in this work is formulated in the spirit of a Gedanken experiment, intended to probe the incompatibility of flavor measurements rather than to realize a loophole-free test of spacetime non-locality. Measurement settings correspond to distinct SM interaction-defined operators -- mass identification, flavor change, and charged-current weak mixing -- that act as alternative measurement contexts on an entangled two-particle quantum state. Within this idealized framework, violations of Bell-type inequalities certify the necessity of entanglement and contextuality in the flavor sector, while remaining agnostic about experimental limitations.

Whether or not entanglement ultimately plays a fundamental role, asking these questions pushes us to re-examine what we mean by measurement, observables, and quantum structure in particle physics.

\begin{acknowledgments}
 {\bf Acknowledgments:} We thank Carlos Wagner for useful discussions. NRS thanks the organizers and participants of the CORFU2025 workshop on the Standard Model and Beyond where some of these ideas were initially presented. NRS is supported by, and CP is partially supported by the  U.S. Department of Energy under Contract No. DESC0007983.
 \end{acknowledgments}

\bibliography{BellsBib.bib}

\end{document}